\documentclass[times,twocolumn,final,authoryear]{elsarticle}
\usepackage{jasr}
\usepackage{framed,multirow}

\usepackage{amssymb}
\usepackage{amsmath}
\usepackage{latexsym}
\usepackage{graphicx}
\usepackage{epstopdf}
\usepackage{color}

\usepackage{amssymb}

\usepackage{anyfontsize}

\newcommand{\lsim}{\lesssim\!}
\newcommand{\gsim}{\gtrsim\!}

\newcommand{\Msun}{M_\odot}

\newcommand{\vect}[1]{\boldsymbol{#1}}

\newcommand{\MsunYr}{\,M$_{\odot}$\,yr$^{-1}$\,}
\newcommand{\kms}{km s$^{-1}$}

\usepackage{url}
\usepackage{xcolor}
\definecolor{newcolor}{rgb}{.8,.349,.1}

\usepackage[citebordercolor=white]{hyperref}

\journal{Advances in Space Research}

\begin{document}

\verso{M. E. Kalyashova \textit{etal}}

\begin{frontmatter}
\title{Direct simulations of very high energy cosmic ray acceleration in 3D MHD model of a compact star cluster}%
\author[1]{M.E. {Kalyashova}\corref{cor1}}
\ead{m.kalyashova@gmail.com}
\author[1]{A.M. {Bykov}}
\cortext[cor1]{Corresponding author}
  \ead{byk@astro.ioffe.ru}
\author[1]{D.V. {Badmaev}}
  \ead{danirbadmaev.astro@gmail.com}
\address[1]{Ioffe Institute, Saint Petersburg, Polytechnicheskaya str., 26, 194021, Russia}
%\address[2]{Affiliation 2, Address, City and Postal Code, Country}
\received{}
\finalform{}
\accepted{}
\availableonline{© 2026. This manuscript version is made available under the CC-BY-NC-ND 4.0 license \url{https://creativecommons.org/licenses/by-nc-nd/4.0/}}
%\availableonline{}
\communicated{}
\begin{abstract}
%%%
Young compact clusters of massive stars contain dozens of O-, B- and WR-type stars with fast powerful winds in a small $\sim$ pc radius core. The particle acceleration by ensembles of shocks accompanied with long-wavelength compressions and rarefactions in the turbulent environment of young massive star clusters (YMSCs) is an alternative to the standard paradigm of Galactic cosmic ray acceleration on supernova shocks. In recent years, the topic has been of great interest due to the fact that modern gamma- and X-ray observatories have detected the radiation from YMSCs, which indicates particle acceleration processes in these objects. We study particle propagation and acceleration in a YMSC with the help of 3D magnetohydrodynamic (MHD) modeling using an open source PLUTO code. The code allows modeling of the turbulent environment of YMSCs and obtaining crucial for particle acceleration values of velocity, density, and magnetic field inside the cluster core. The Particle module implemented in PLUTO allows solving  the equations of motion for test charged particles together with MHD equations for the medium. We obtained that protons acceleration up to hundreds of TeV takes place in the cluster core near the termination shocks of O-stars, which are surrounded by shocks of their neighbour stars. We also modeled an interesting case of a young supernova remnant expanding inside the cluster core. In this case a very fast acceleration takes place: particle energies $\gsim$ 100 TeV can be obtained in $\lsim$ 100 years. The particle spectra and spatial distribution are discussed.  
%%%%
\end{abstract}
\begin{keyword}
%% Keywords
\KWD cosmic rays, particle acceleration, MHD, star clusters, stellar winds, supernova remnants
\end{keyword}
\end{frontmatter}

\makeatletter
\let\savedefaultfootnote\footnote
\let\savedefaultfootnotetext\footnotetext
\makeatother

%% For linenumbers
%\linenumbers

\section{Introduction}

The search for sources of Galactic cosmic rays, including those with energies higher than petaelectronvolts (PeV), is one of the long-term problems of high-energy astrophysics. Inside young massive star clusters (YMSCs) huge mechanical energy is released in the form of supernovae and powerful stellar winds. Some of this energy is converted into the acceleration of the particles, which can gain energy due to the collective plasma effects \citep[][]{Bykov2001}. In recent years, the topic has attracted significant interest because of the results of modern gamma-ray  telescopes. The Fermi orbital observatory, as well as ground-based telescopes HESS, HAWC, MAGIC, LHAASO, Tibet,  Carpet-2 detect high gamma-ray fluxes in the direction of known star clusters \citep[][]{AharonianNat2019, YangWd2}. Many gamma-ray sources have been associated with clusters: Westerlund 1 \citep[][]{HessWd1} and Westerlund 2 \citep[][]{YangWd2}; RSGC 1 \citep[][]{RSGC12020}; NGC 3603 \citep[][]{NGC3063}; Cygnus Cocoon, in the region of which there are several stellar associations, the largest is Cygnus OB2  \citep[][]{AckermannSB2011, HonaNatAs2021, LhaasoCyg2024}; 30 Doradus C in the Large Magellanic Cloud \citep[][]{LMC2015} and others. Some massive clusters are also bright X-ray sources: recently  \citet{BykovWd2} showed that the X-ray emission of the Westerlund 2 cluster has a non-thermal component, which clearly indicates particle acceleration processes in the cluster core.

The question of whether massive stellar associations and clusters can efficiently accelerate cosmic rays (CRs) is tested by examining the spatial and spectral characteristics of $\gamma$-ray emission from Galactic OB associations and compact clusters. Using Fermi-LAT and H.E.S.S. data for the Cygnus OB2 association and the compact clusters Westerlund 1 and Westerlund 2, \citet{AharonianNat2019} analyzed the spatial distribution of $\gamma$-ray emission from these regions. They found that the CR density follows a radial dependence of $r^{-1}$, concluding that relativistic particles are continuously injected into the interstellar medium. 
Although the $r^{-1}$ profile may be due to the different reasons, e.g. the self-confinement of accelerated CRs, it also may support the hypothesis that massive stellar clusters and OB associations are probable Galactic CR accelerators.

Gaia DR2 survey identifies $\sim$ 2000 stellar clusters located in the Galactic disc \citep[][]{CantatGaudin2020}. From \citet[][]{Portegies_Zwart2010} about a dozen of them has mass $> 10^4 ~ M_{\odot}$ and age $< 20$ Myr.  The total kinetic power of massive stellar winds in the Galaxy is estimated as $\sim 10^{41}$ erg~s$^{-1}$ \citep[see][]{Seo2018}, while it is expected that at least 50\% of this power can be related to clusters and associations. Such kinetic power is unsufficient to provide the observed GeV CR flux, but at TeV-PeV energies clusters can contribute significantly. Both superbubbles and massive star clusters are probably dominant sources of very-high-energy $\gamma$-rays in starburst galaxies \citep[][]{Ohm16}.

In colliding flow systems, like clusters of massive stars, there can be a significant enhancement of the turbulent magnetic field, and part of the mechanical energy of the winds or supernova shocks converts into magnetic energy. An intermittent highly turbulent medium with strong primary and weak secondary shocks is formed with a wide range of MHD fluctuations. 
Particle acceleration by ensembles of shocks and waves of compression and rarefaction in the turbulent environment of star clusters may be an alternative to the standard paradigm of the acceleration of Galactic cosmic rays on the shock waves of supernova remnants. The energy release of massive clusters allows particle acceleration up to hundreds of TeV and beyond. 

A number of models of particle propagation and acceleration in turbulent environments of massive clusters and superbubbles were proposed  \citep[][]{Bykov2001,BT2001, Parizot2004, Ferrand2010}. Most of them are at least partially analytical. In particular, in \citet{Bykov2001} an analytical model for particle propagation in turbulent plasma with shock fronts was developed based on the renormalization method. The essence of the model lies in solving the particle transport equation averaged over an ensemble of turbulent motions. 
A similar approach is developed in works \cite{Vieu22a, Vieu22b}, devoted to particle acceleration in superbubbles, formed by the collective action of stellar winds in OB associations. The soft spectra of the accelerated particles and their maximum energies $\lsim 0.5$ PeV are predicted. In other works, such as \citet{Morlino21}, it is assumed that particle acceleration in OB associations and compact clusters takes place on a termination shock of a collective cluster wind.

While these approaches provide valuable insights, a more precise method for studying turbulent astrophysical environments is the MHD simulation of a complex configuration of plasma flows. For this purpose, we used the open-source code PLUTO \citep[][]{Mignone07,Mig18}, designed specifically for solving astrophysics problems.
Previously, as part of a study of turbulence in YMSCs, 3D MHD simulations of multiple stellar winds in the cluster core were performed \citep[][]{Bad22}. It was shown that, as predicted by the analytical model, magnetic fields in such a system are highly intermittent and amplified up to $\sim$ 300 $\mu$G due to the energy of powerful winds of young stars. In the subsequent work \citep[][]{Bad24}, a supernova in the core of YMSC was simulated and perturbations of density, velocity, and magnetic field as a result of the explosion were studied. The relaxation time of the cluster to its pre-supernova state was obtained, which turned out to be $\sim$ 10 kyr. Thus, through MHD modeling, all the main characteristics of the core of a young cluster, their temporal evolution and the impact of the supernova on the cluster medium were investigated.

The obtained results of MHD simulations can be applied to the problem of particle acceleration. The sizes of the high magnitude magnetic field filaments were found to be $l\sim 0.5$\,pc. One can  simply estimate the maximum energy $\epsilon_m$ of protons accelerated in the system with the Hillas criterion \citep{Hillas84} as 
\begin{equation}
  \epsilon_m \sim \frac{u}{c} elB,  
\end{equation}
 where the plasma bulk velocity field $u$ reaches $\sim 1000~\rm{km~s^{-1}}$. From this estimation it follows that clusters can confine and accelerate protons to energies above $\gsim 100$\,TeV. The highly amplified magnetic fields are surrounding the regions with the strong termination shocks of the fast stellar winds which can inject and accelerate non-thermal particles.

We now aim to move from estimates to modeling and study the propagation and acceleration of cosmic ray particles in the turbulent environment of a star cluster through direct simulation with the PLUTO code. This is possible due to the Particle module implemented in PLUTO, designed to solve the equations of motion of test charged particles together with MHD equations for the medium.
 
In this paper, we focus on young compact clusters of massive stars. In such clusters, dozens of massive O- and WR-type stars are concentrated within a region of $\sim$ several parsecs. The kinetic power of these objects can reach $10^{38}-10^{39}$ erg~s$^{-1}$. Although the maximum energy of particles accelerated in supernova remnants (SNRs) does not exceed tens of TeV, we expect that the maximum energy of accelerated particles in YMSCs may reach hundreds of TeV.

The paper is organized as follows. The stellar cluster setup in the PLUTO code and the details of the particle propagation modeling in a cluster are provided in Section 2. In Section 3 we provide the characteristics of the simulated turbulent medium, discuss the maximum energies obtained and the acceleration sites. In Section 4 we describe the modeling of particle acceleration in a very interesting case of a young supernova remnant going through the cluster core. Finally, in Section 5 we briefly discuss the impact of the dense magnetized shell near the cluster.

\section{Model}

The open source magnetohydrodynamic (MHD) code PLUTO \citep[][]{Mignone07,Mig18,Vai18} is a widely used tool for astrophysical plasma modeling. The code numerically solves the MHD equations using Godunov's scheme. It is used to study high-energy astrophysical systems, such as stellar winds, supernova remnants, pulsar wind nebulae, stellar clusters, and other objects. MHD simulations in PLUTO provide the structure and magnitudes of magnetic fields, density, pressure, and velocity distributions in regions of interacting winds from massive stars. 

The simulation was conducted in two stages. In the first stage, the environment for particle propagation was prepared through 3D MHD modeling of the collision of several dozen stellar winds from massive stars concentrated in the core of a YMSC. After a period of evolution, the system reached a quasi-stationary state, in which the spatial configuration of the flows no longer changed significantly over time.
The second stage focused on particle propagation and acceleration. To conserve computational resources, a "snapshot" of the quasi-stationary cluster environment was taken. Particle trajectories were then calculated within this static snapshot, without solving the MHD equations. This approach enabled particle propagation to be extended to several thousand years.

\subsection{Modeling the cluster} \label{sec:clust}
The thorough description of the modeling of the YMSC core with the PLUTO code is provided in \citep[][]{Bad22}. The code integrates the following set of non-ideal magnetohydrodynamic equations: 
\begin{gather}
    \frac{\partial\rho}{\partial{t}}+\vect{\nabla}\cdot\left(\rho\vect{u}\right)=0,\label{1}\\
    \frac{\partial\vect{m}}{\partial{t}}+\vect{\nabla}\cdot\left(\vect{m}\otimes\vect{u}-\vect{B}\otimes\vect{B}+\vect{\hat{I}}p_{\mathrm{t}}\right)=0,\label{2}\\
    \frac{\partial{E}}{\partial{t}}+\vect{\nabla}\cdot\left[\left(E+p_{\mathrm{t}}\right)\vect{u}-\vect{B}\left(\vect{u}\cdot\vect{B}\right)\right]=\Phi\left(T,\rho\right),\label{3}\\
    \frac{\partial\vect{B}}{\partial{t}}+\vect{\nabla}\cdot\left(\vect{u}\otimes\vect{B}-\vect{B}\otimes\vect{u}\right)=0,\label{4}
\end{gather}
where $\vect{m}=\rho\vect{u}$ represents the momentum density vector of a control volume, $\vect{B}$ is the magnetic field vector (which is normalized by $\hspace{-3pt}\sqrt{4\pi}$ in the code), $\vect{\hat{I}}$ is the identity matrix, and $p_{\mathrm{t}}=p+\vect{B}\cdot\vect{B}/2$ is the total pressure. The total energy density reads,
\begin{equation}
    E=\frac{p}{\gamma-1}+\frac{\vect{m}\cdot\vect{m}}{2\rho}+\frac{\vect{B}\cdot\vect{B}}{2},
\end{equation}
where $\gamma=5/3$ is the ratio of specific heats for a monoatomic ideal gas. We take into account the energy gains and losses by optically thin radiative processes, $\Phi\left(T,\rho\right)$, following the recipe from \citet[][]{Mey14} for the case of a photoionized medium. The temperature is obtained from the ideal gas law,
\begin{equation}
T=\mu\frac{m_{\rm{H}}}{k_{\rm{B}}}\frac{p}{\rho},
\end{equation}
with the mean mass per particle, $\mu=0.61$, for gas ionized by the radiation of hot massive stars, and the mass of a hydrogen atom $m_{\rm{H}}$. The sound speed of the plasma, $c_{\mathrm{s}}=\sqrt{\gamma{p}/\rho}$, closes the system of equations described above.

The adopted numerical scheme contains a finite-volume, dimensionally unsplit Godunov-type solver made of the combination of HLLD \citep[][]{MK05} and HLL \citep[][]{HLL83} Riemann solvers for flux computation; a $2^{\rm{nd}}$-order linear (TVD) reconstruction for spatial accuracy; a van Leer slope limiter for flux limiting; a $2^{\rm{nd}}$-order Runge-Kutta (RK2) integrator for time stepping; and the standard Courant-Friedrichs-Lewy condition, which is set to $C_{\mathrm{cfl}}=0.2$, to determine the time step value. Throughout the simulation, the divergence-free condition for the magnetic field, $\vect{\nabla}\cdot\vect{B}=0$, is ensured by the Hyperbolic Divergence Cleaning algorithm \citep[][]{Ded02}.

The simulation setup includes two types of massive stars: O-type (with mass loss $\dot{M} = 10^{-6}~ \rm{\Msun ~ yr^{-1}} $ and wind velocity $v_w = 2300 ~\rm{km ~s^{-1}}$) and Wolf-Rayet ($\dot{M} = 10^{-5}~ \rm{\Msun ~ yr^{-1}}$, $v_w = 1800 ~\rm{km ~s^{-1}}$).
All massive stars are concentrated in the cluster core of 2 pc radius, while the total computational domain is a cube with a side of 12 pc. This size was chosen in order to suppress particle escape and to take into account the cluster surroundings. We employed different spatial resolutions for the cluster core ($0.013 ~\rm{pc~cell^{-1}}$ in the central domain of $4 \times 4 \times 4 ~\rm{pc}^3$) and for the surroundings ($0.04 ~\rm{pc~cell^{-1}}$) in order to better resolve the interacting winds and stellar termination shocks. The total number of cells is $500^3$.

Massive O- and WR-type stars exhibit different wind velocities, rotation speeds, and mass-loss rates. To evaluate how stellar composition affects cluster properties and particle acceleration, two simulations with identical total wind kinetic power ($10^{38}$ erg~s$^{-1}$) but different stellar populations were performed:
\begin{itemize}
       \item 12 O-stars + 8 WR-stars (case 1)
       \item 48 O-stars + 2 WR-stars (case 2)
\end{itemize}

We ran the simulation up to the moment when the stellar wind turbulence fills the entire volume and the configuration of flows becomes quasi-stationary. After this moment, the structure of flows and the main characteristics in the domain experience only minor changes. The time required for this is $\sim 25000~\rm{years}$.

\subsection{Modeling the particle propagation}\label{sec:clustprop}
The second part of the modeling was the simulation of the particle propagation in the "snapshot" of a stellar cluster obtained in \ref{sec:clust}. It was performed using the Particle module in the PARTICLE$\_$CR regime, which provides the direct solution of equations of motion for test particles in the medium \citep[][]{Mig18,Vai18}. The MHD equations are not solved during this particle propagation stage in order to speed up the simulation under the assumption that on the scales of $\sim1000$ years the flows are stationary. The particle injection takes place at the fronts of strong termination shocks (Mach number $\gsim100$) of massive stellar winds in the cluster core. A total of $10^5$ particles (protons) are simultaneously injected with an initial energy of 30 TeV (monoenergetic injection). The choice of such high injection energy is due to the simulation timestep, which increases proportionally to the particle energy, speeding up the simulation. For the same reason if a particle's energy drops below a threshold $E_{\rm{min}}$ (which is $5\times 10^{12}-10^{13}$ eV in different simulations), it is removed from the simulation. We are aware that to restore a realistic particle spectrum, one should be more careful with the assumptions about injection energies and spectra. However, our approach is reasonable for the purpose of finding the maximum energy attainable in a cluster.

\section{Results}

\subsection{Magnetic fields}
\begin{figure}[h]
  \centering
  \includegraphics [width=85mm] {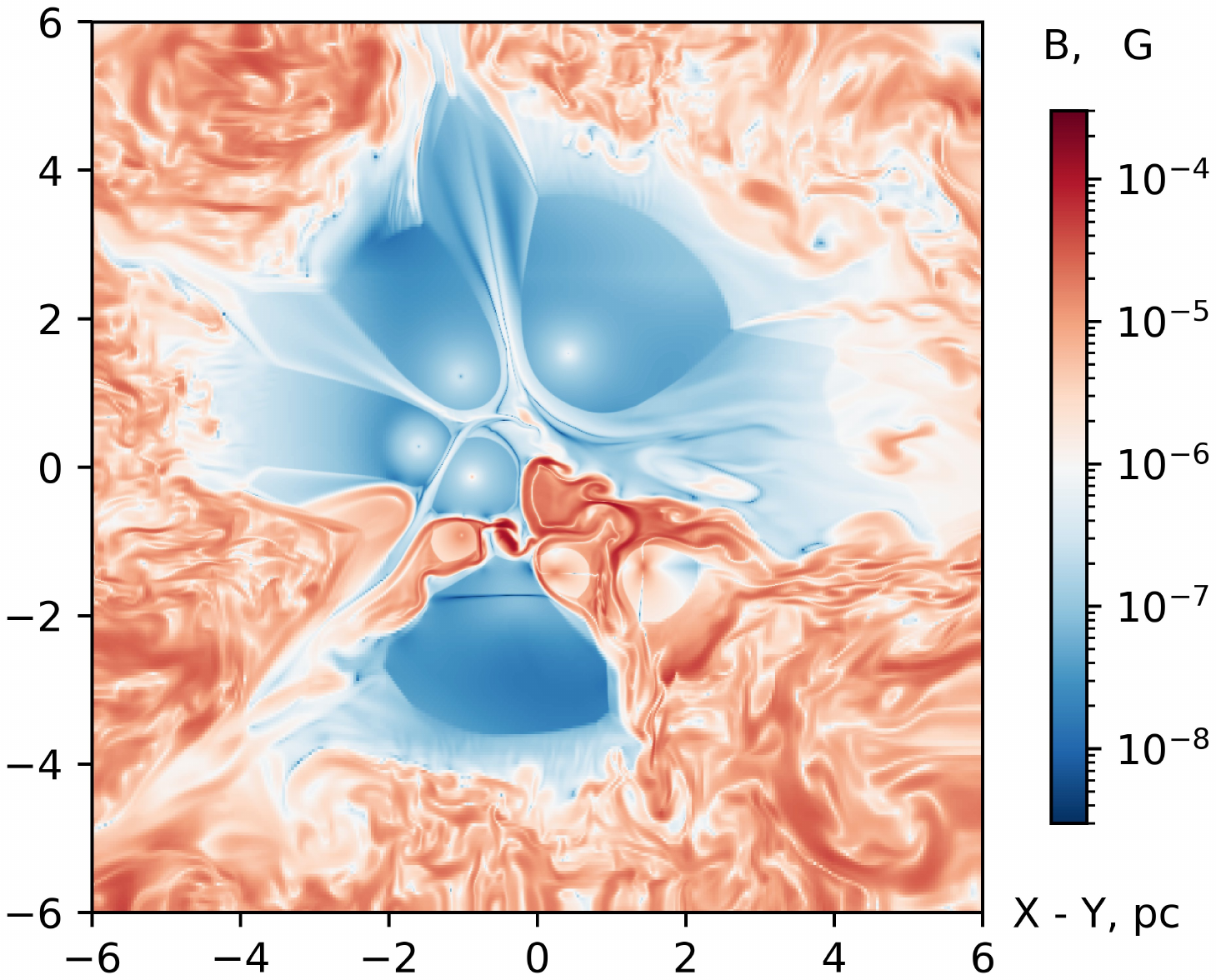}
  \includegraphics [width=85mm] {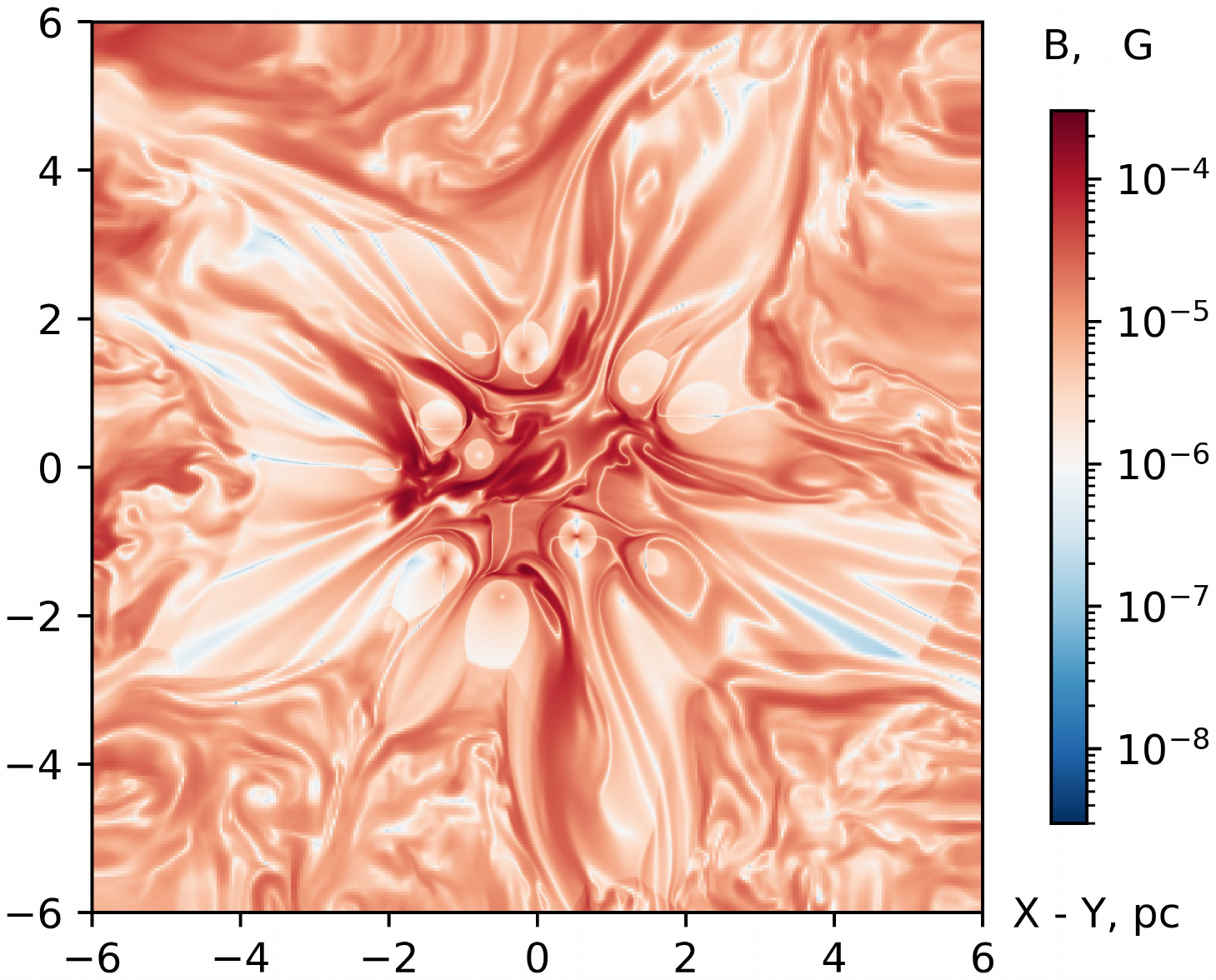}
  \caption{Magnetic field map of the central Oxy-plane of the compact star cluster and its surroundings for the case (1) (12 O-stars + 8 WR-stars) – upper panel, and (2) (48 O-stars + 2 WR-stars) – lower panel.}
  \label{mfields}
\end{figure}

Figure~\ref{mfields} reveals the complex morphology of magnetized plasma flows within the cluster, shaped by the interaction of multiple winds from hot massive stars. Due to the high mass loss rate ($\dot{M}=10^{-5}$\MsunYr), a single WR-wind releases nearly 6 times more mechanical power compared to an O-wind; this results in much broader wind-blown bubbles associated with WR-stars, which dominate the cluster core volume in the case (1) and are clearly visible as low-field blue cavities extended outwards. At the cluster periphery, behind the termination shocks of WR-winds, a large-scale turbulent swell with amplified magnetic fields is formed. Highly magnetized bow shock structures are concentrated around less powerful O-winds.

In case (2) higher mean magnetic fields are produced, though the maximum field strengths in both cases are $\sim $300 $\mu$G. The reasons for this are the higher number of stellar wind sources as well as the higher rotation speed of O-stars compared to WRs \citep[e.g.][]{MM03,Eks12}, which affects the dominant azimuthal, $B_\phi$, component of the stellar magnetic field. In contrast, the lower panel of Figure~\ref{mfields} displays a more filamentary and compact magnetic field structure. The reduced number of WR winds, replaced by a larger number of weaker O-star winds, leads to more centrally localized turbulence and enhanced field compression along radial outflows. The presence of dense radial filaments suggests efficient magnetic field amplification through wind-wind interactions. 

As the wind velocities of O- and WR-stars are comparable, the resulting mean flow velocities in the domain do not show significant differences. The case comparison highlights how varying the stellar composition – while keeping the total wind kinetic power constant – significantly alters the magnetic field configuration within the cluster, influencing plasma dynamics.

\subsection{Maximum energies and spectra}

We performed the particle propagation simulations in the cluster medium for cases (1) and (2). The rates of the particle acceleration in cases (1) and (2) do not show significant differences. For example, during the first 1500 years of calculation we got the maximum energies $\sim 330$ TeV for case (1) and $\sim 320$ TeV for case (2). 

For the full simulation durations we obtained the following maximum energies of accelerated particles:

(1) $\sim$390 TeV (for $\sim$4000 yrs of particle propagation)

(2) $\sim$340 TeV (for $\sim$2000 yrs of particle propagation)

Particles reaching such energies form the high-energy tail of the resulting energy distribution.

\begin{figure}[ht]
\centering
\includegraphics [width=85mm] {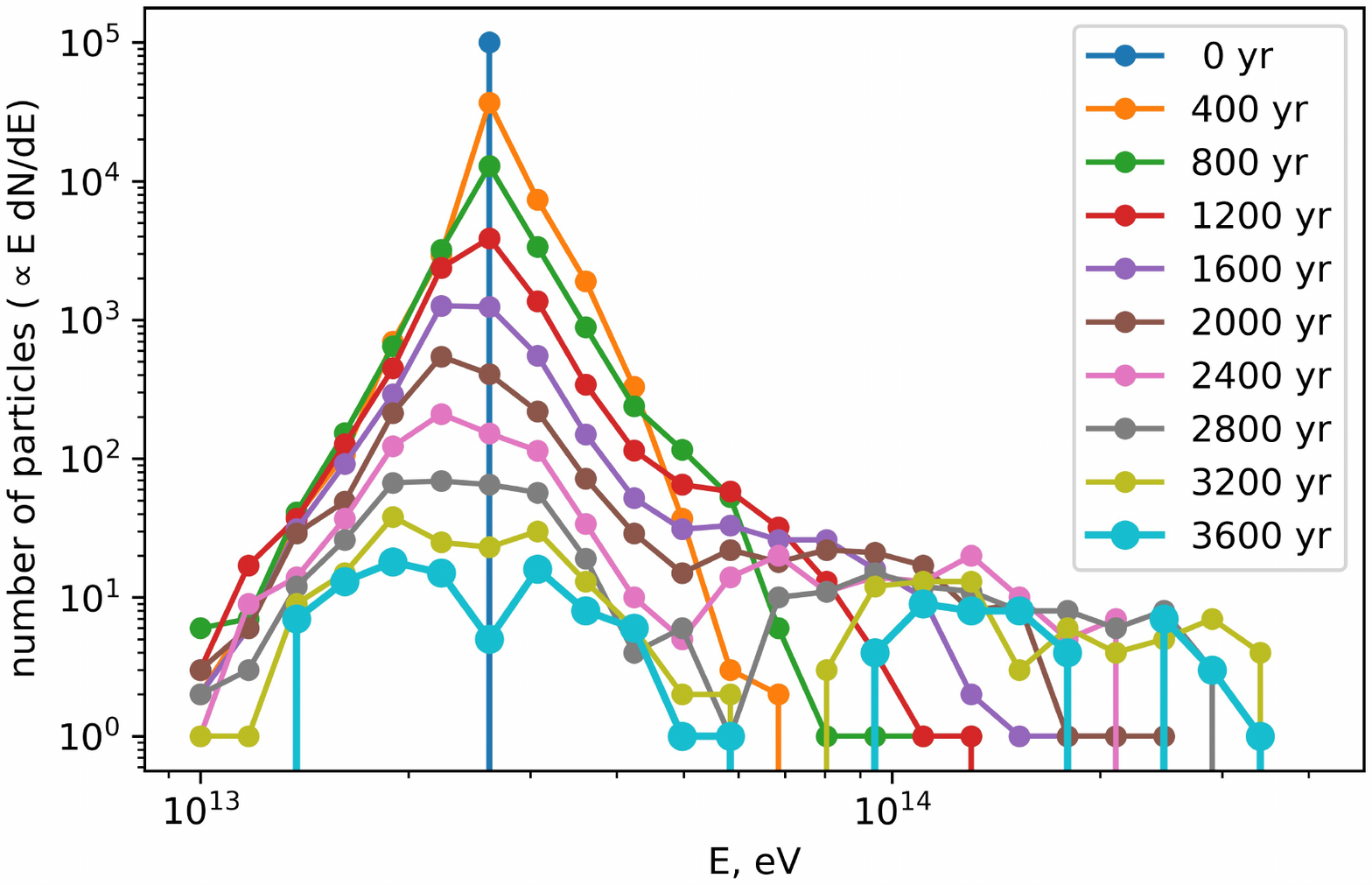}
\caption{Time dependence of the spectrum of accelerated particles for the case (1). Particles were injected at $t=0$ with energy $E=30~\rm{TeV}$. As the spectral bin width $\propto E$,  the number of particles  $\propto E ~dN/dE$.}
\label{spec1}
\end{figure}

For case (1), the dependence of the number of accelerated particles on energy is presented in Fig.~\ref{spec1} for times from $t=0$ (monoenergetic impulsive injection at 30 TeV) to $t=3600$ years. As the spectral bin width $\propto E$,  the number of particles in this Figure $\propto E ~dN/dE$. One can see from the Figure, that the distribution becomes much flatter with time and at times $>3000$ years the numbers of particles remaining in the domain with low and high energies are of the same order.

\begin{figure}[ht]
\centering
\includegraphics [width=85mm] {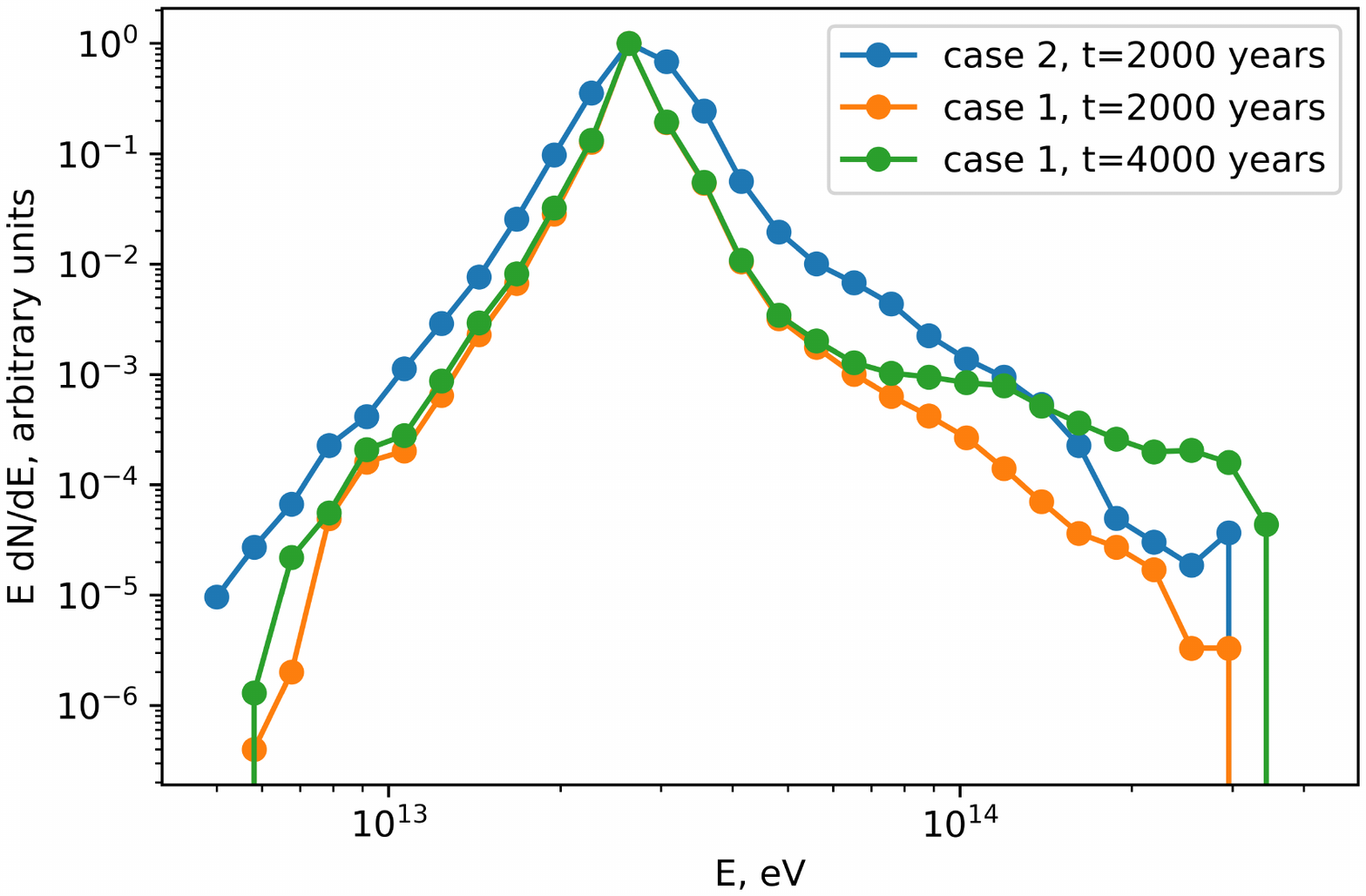}
\caption{Time-integrated spectrum of accelerated particles, normalized on the maximum value.}
\label{spec_int}
\end{figure} %КАРТИНКА МОЖЕТ УБИВАТЬ КОМПИЛЯЦИЮ
 
Given that we employed impulsive particle injection, the time-dependent spectrum shown in Fig. \ref{spec1} is the analogue of the time-dependent Green's function $G(E, t)$ for particle acceleration and escape in this system. To obtain the spectrum for a scenario with continuous injection over time, we convolved this Green's function with a constant injection rate by integrating our time-dependent solution over the simulation period $[0, T_{sim}]$. The integrated spectra  are presented in Figure \ref{spec_int}. One can see, comparing the distributions for case (1) at $t=2000$ years and $t=4000$ years that the spectrum gets harder at the highest energies, and a continuing upward trend may be expected. However, we cannot definitively verify this behaviour as at $t=4000$ years nearly all particles have already escaped the system and the further integration is impossible. We also note that the spectral shape for the case (1) and (2) is quite similar.

As the injection is monoenergetic, we can not derive the impact of the injection spectrum on the final spectral distribution. We assume that the initial spectral shape would dominate with time just like in Fig.\ref{spec_int} one can see the injection peak. Due to the acceleration and diffusion in momentum space, we also would see high-energy and low-energy tails, slowly rising with time. We expect that the high-energy tail of the distribution would remain similar as in the present calculation.

\subsection{Particle escape and diffusion}

\begin{figure}[ht]
\centering
\includegraphics [width=85mm] {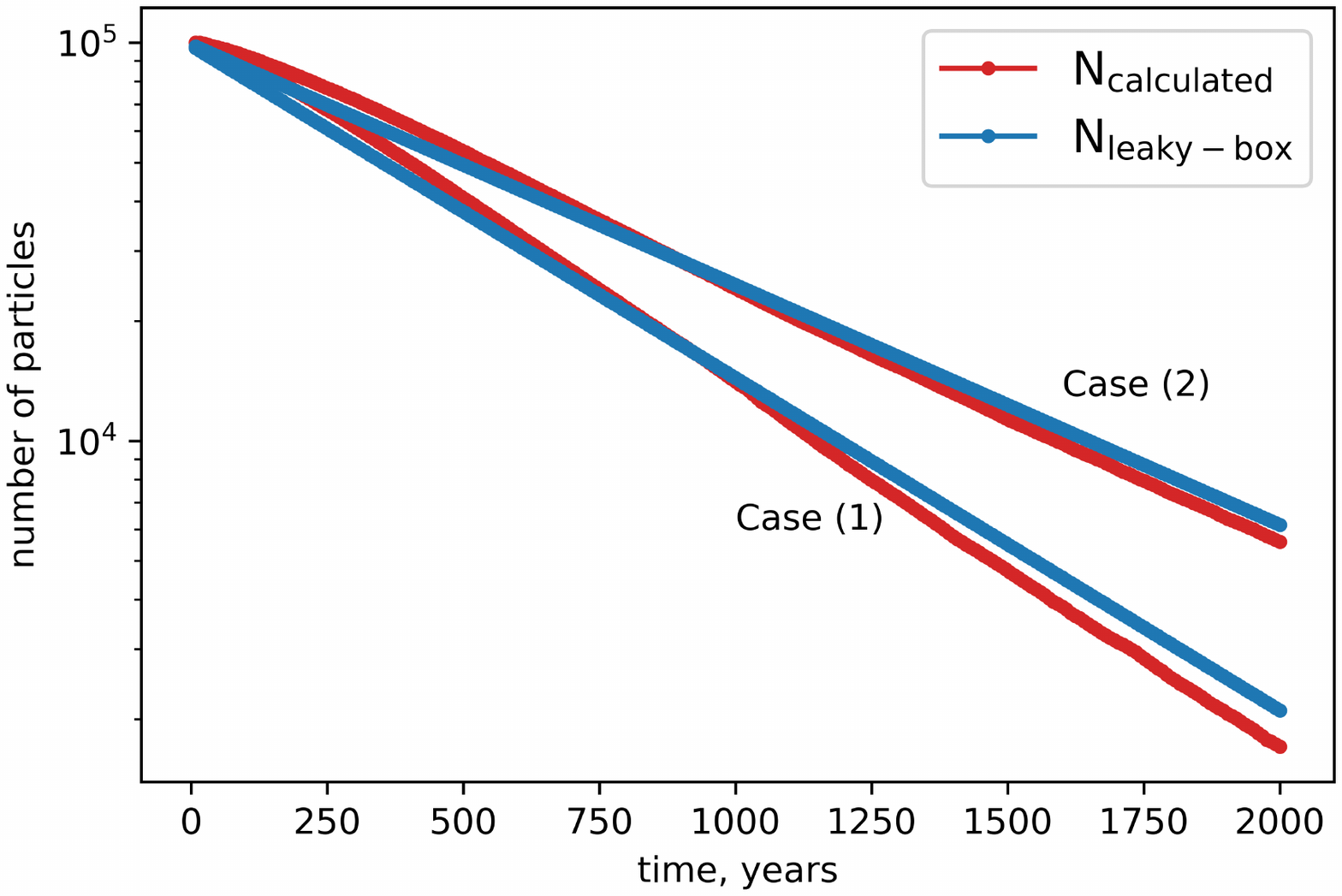}
\caption{
Number of particles in the domain vs. time in simulation (red line) and leaky box model (blue line) for case (1) (upper panel) and case (2) (lower panel)}
\label{esc_leaky}
\end{figure}

To estimate the rate at which particles leave the computational domain, one can use the leaky-box model, which predicts the following dependence of the number of particles on time, given an impulsive injection:
\begin{equation}
    N=N_0 e^{-t/T_{esc}}
\end{equation}
 
 Fig. \ref{esc_leaky} shows that this model provides a good approximation of the particle escape from the cluster. The values of $T_{esc}$ then are
\begin{equation}
T_{esc} \approx 520 ~\rm{years ~(case ~1)}
\end{equation}
\begin{equation}
T_{esc} \approx 720 ~\rm{years ~(case ~2)}  
\end{equation}
 
As expected, the stronger magnetic fields in case (2) make it harder for particles to escape.

Assuming the characteristic size of the computational domain is $R \approx 6~ \rm{pc}$, the corresponding to $T_{esc}$ spatial diffusion coefficient can be estimated as  $D \simeq R^2/6T_{esc} \approx 3 \times 10^{27}$ $\rm{cm^2~s^{-1}}$.  This value is average  for the entire particle sample of all energies. It is suppressed by a factor of $\sim 100$ with respect to the average Galactic diffusion coefficient for energies of tens of TeV, estimated from B/C constraints \citep[see, e.g.,][]{Giacinti_2018}.

It is important to keep in mind that the model does not consider the magnetic field amplification via cosmic ray driven instabilities and the presence of small-scale (on scales smaller than the simulation resolution) turbulence. Additional turbulence can slow down the particle escape.

\subsection{Sites of particle acceleration}

\begin{figure}[h]
\centering
\includegraphics [width=85mm] {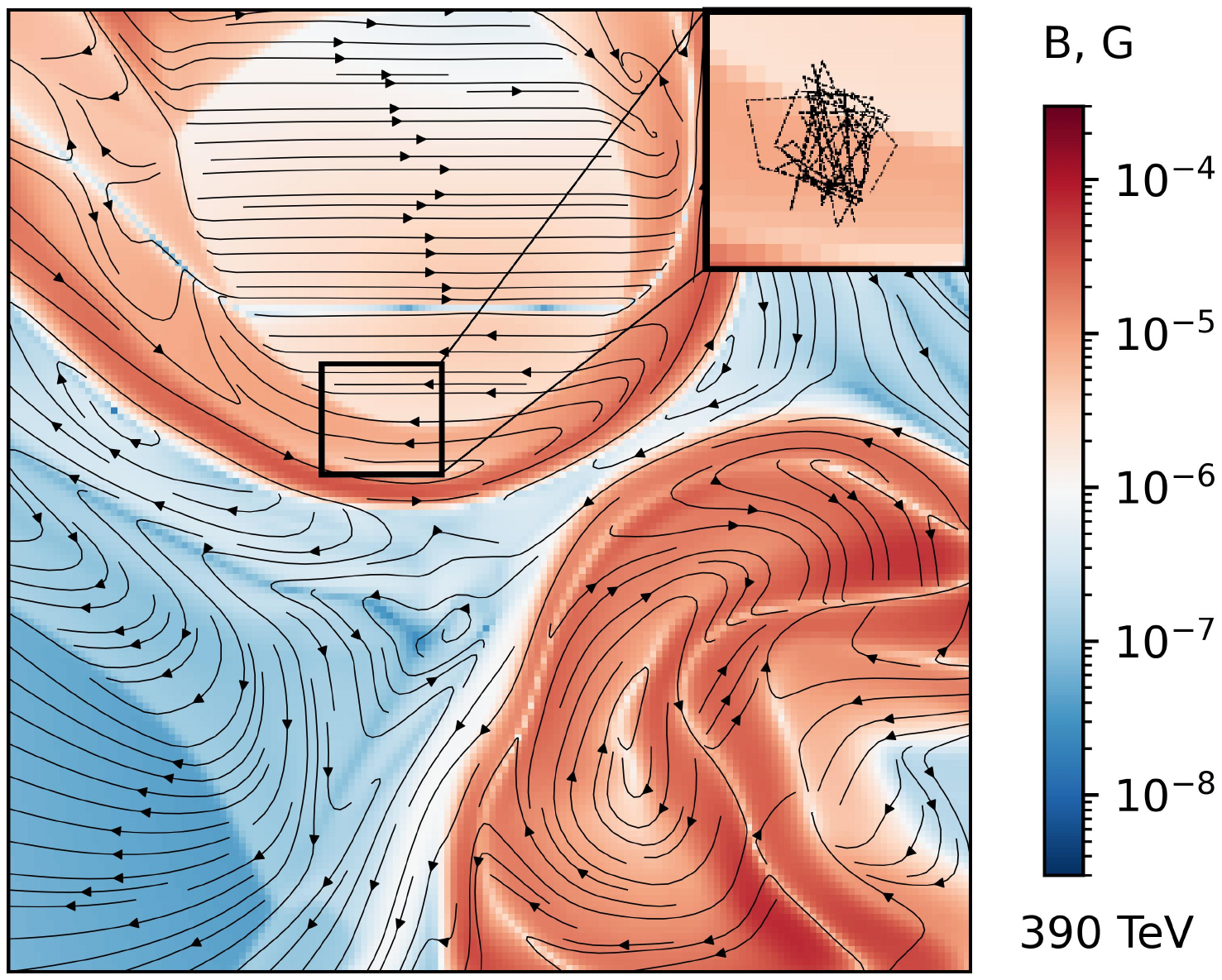}
\includegraphics [width=85mm] {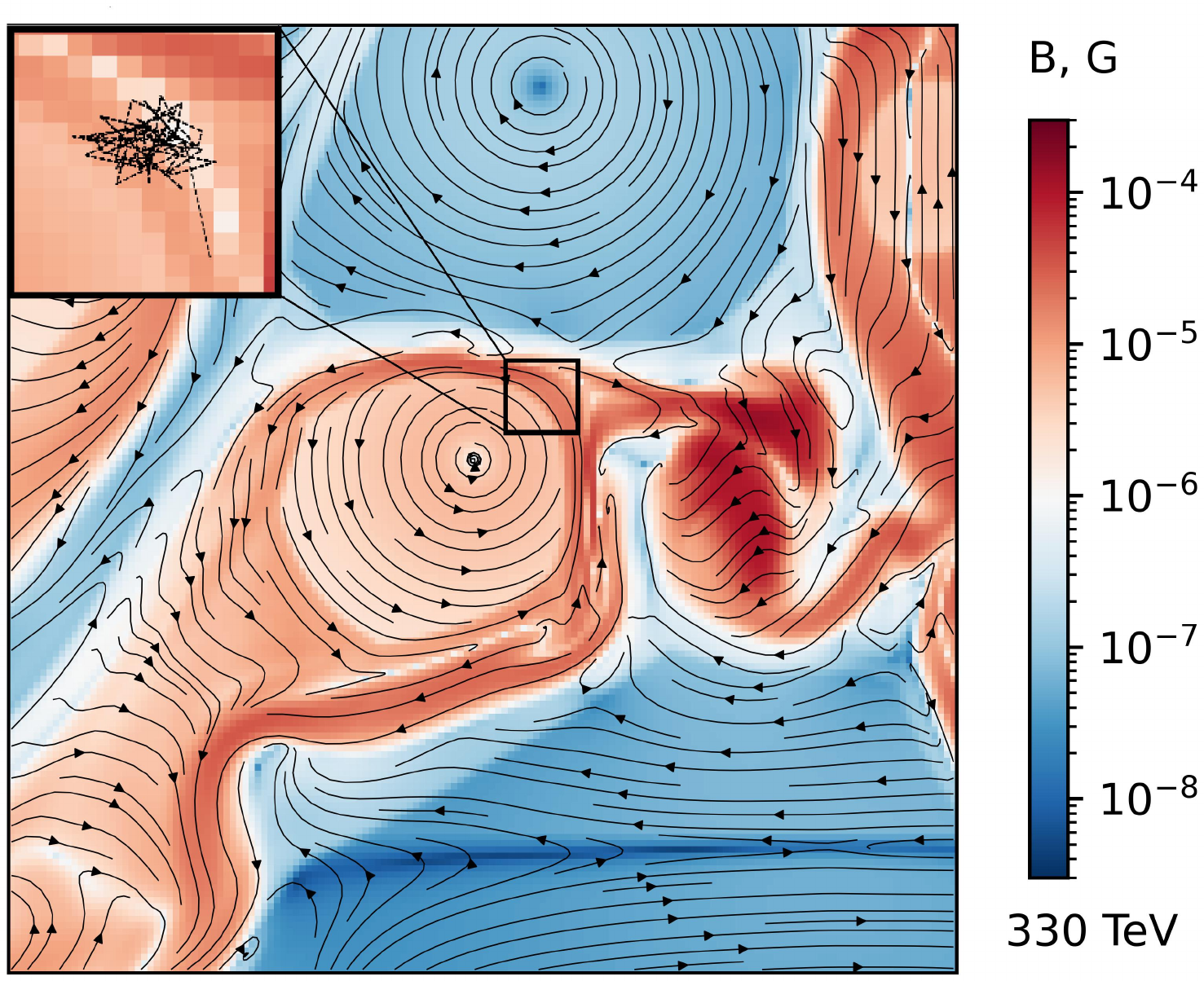}
\caption{Examples of the energy gain regions of particles accelerated to $>$ 300 TeV. The magnetic field geometry is shown with streamlines. Insets show the trajectories of accelerated particles with energies 390 TeV (upper figure) and 330 TeV (lower figure)  during the acceleration process.
}
\label{traj}
\end{figure}

The acceleration mechanism in the cluster combines acceleration at the stellar wind termination shocks and stochastic acceleration via scattering on magnetic turbulence generated in the cluster. To locate the dominant regions of acceleration, trajectories of the highest-energy particles ($>$100 TeV) were tracked. The particles were found to gain most of their energy near the shock fronts where they were injected (see Fig. \ref{traj}). Once a particle leaves the vicinity of a shock, it ceases to gain energy.
This can explain naturally the fact that despite the different escape times for particles in cases (1) and (2), the maximum energies reached are quite similar: the acceleration regions are local and the particle escape from the acceleration sites is regulated by the amplified magnetic field between the shocks, which can reach hundreds of $\mu G$. From Fig. \ref{spec1} it follows that some particles from the high-energy tail are trapped in their acceleration sites and stay there for a long time, gaining energy. In the absence of CR driven small-scale turbulence, needed for the effective particles scattering in diffusive shock acceleration, the probable mechanism is shock drift acceleration. Large-scale turbulent magnetic structures and possible magnetic mirrors\footnote{For a recent discussion of magnetic mirrors in cosmic ray transport see \citet{2021ApJ...923...53L}, \citet{2022PhRvL.129u5101L} and \citet{2025MNRAS.539.1236B}} are able to confine particles near the shock, allowing them to engage in a shock drift acceleration process multiple times.

At the same time, one can see from Figs. \ref{spec1}, \ref{spec_int} that particles can not only gain, but also lose energy, which implies the presence of Fermi second-order acceleration, occurring on long-wavelength compression-rarefaction fluctuations in the cluster surroundings. Besides direct numerical calculation presented here, this mechanism can be described based on Fokker-Planck type equation for intermittent systems with multiple shocks \citep[see, e.g.,][]{BT93}.

\section{Supernova event in the cluster core}
An efficient acceleration may occur if a supernova explodes within the cluster, due to the high velocity of the supernova ejecta and the formation of a converging shock system as the supernova’s forward shock propagates through the cluster core. 

The SNR shock collides with lots of powerful stellar winds inside the YMSC core, creating a complex shocked environment \citep[][]{Bad24}. This results in a highly perturbed and filamentary structure of the magnetic field that reaches magnitudes $\gtrsim100$ $\mu$G. This highly violent plasma state exists for $\lesssim 1000$ years. 

In systems with converging flows, particle confinement is more efficient than in the vicinity of individual shocks. This modifies the diffusive shock acceleration mechanism and leads to higher maximum particle energies.

Supernova event in the massive cluster core is a promising way to obtain particle energies much higher than expected from the individual SNRs. Nonlinear modeling of the acceleration process in YMSC with SN was performed in \citet{BEGO2015MNRAS} in an attempt to check the possibility of the production of PeV neutrinos from the IceCube map by accelerated protons in Westerlund 1. The mechanism operates only for a short time until the shock of supernova (which occured outside the cluster core)  collides with the collective cluster wind ($\lsim$ 400 years, an estimate derived for Westerlund 1). During this phase, only the highest-energy particles can escape the system. The modeling demonstrated that the accelerated particle spectrum in supernova-wind systems can reach energies up to 40 PeV and exhibits a very hard behavior at high energies. The authors showed that to accelerate CRs to 40 PeV the high supernova shock velocities of $\sim$ 10000 \kms~ and magnetic field magnitudes of $\sim$ 1 mG in the shock vicinity are required. The magnetic field amplification in their approach is due to the cosmic ray driven instabilities and obtained in kinetic Monte Carlo simulation. These demanding conditions are relaxed for PeV range proton acceleration. In 3D MHD modeling of a supernova evolution with 10 $\Msun$  ejecta mass and 10$^{51}$ erg energy in a compact cluster by \citet{Bad24} the velocity of forward shock of about 7000 \kms~ and magnetic field magnitudes up to 0.4 mG were obtained even without an account of magnetic field amplification in diffusive shock acceleration. The conditions achieved in these simulations are favourable for PeV regime CR acceleration. 

We performed similar MHD simulations of a supernova explosion in a compact cluster and examined directly particle acceleration in this system. 
To initialize an SNR within the cluster core, we employed a mapping strategy \citep[e.g.][]{Mey15,Bad24} immediately after the flow configuration of stellar winds became quasi-stationary. At first, we performed one-dimensional (1D) hydrodynamical simulations of the young SNR expanding into the wind of the progenitor star with extremely high resolution. For the 1D SNR model, we follow a standard initialization recipe \citep[e.g.][]{CL89,TM99,Wha08} which leads to the classical self-similar solution \citep[][]{Chv82,Nad85} consisting of a reversed shock, contact discontinuity, and a forward shock. As soon as the 1D SNR shell expands enough to be adequately resolved inside the 3D computational grid (domain) described in Section~\ref{sec:clust}, we isotropically map the obtained 1D profiles of density, pressure, and velocity into the pre-simulated 3D environment of the cluster core, replacing the wind of the progenitor star. After this mapping procedure, we continue the MHD simulation of gas dynamics inside the cluster core.% for another 100-1000 years.

\begin{figure}[ht]
\centering
\includegraphics [width=85mm] {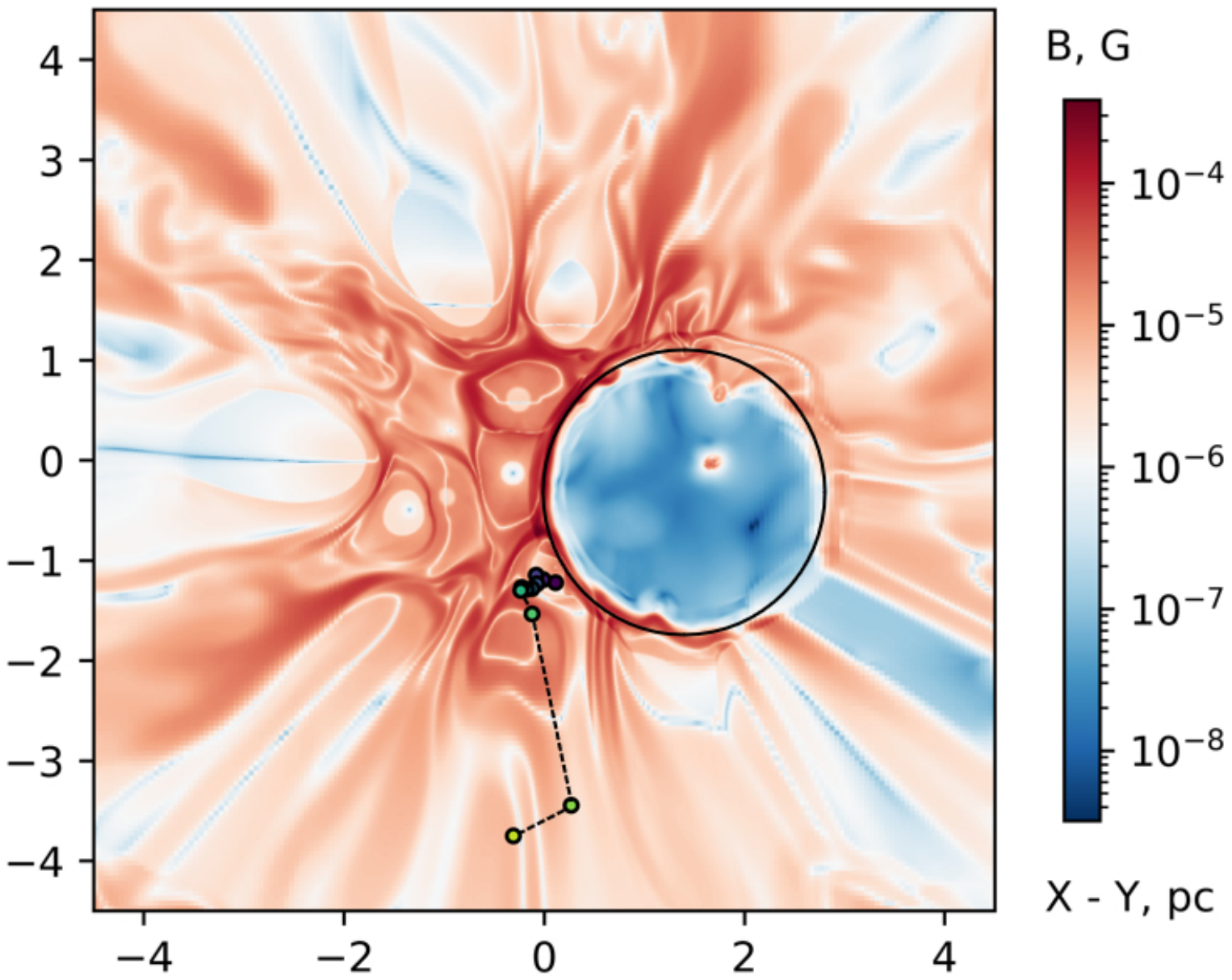}
\caption{Magnetic field map during a SN event in YMSC. An approximate position of SNR shock is indicated with a black circle. An accelerated particle trajectory is shown with a dashed black line.}
\label{sn_mf1}
\end{figure}

In the presence of SNR, the "snapshot" approach could not be used, as the supernova shock rapidly traverses the cluster core. Therefore, the MHD equations have to be solved together with particle equations of motion, which significantly slows down the simulation. This is the reason why the modeling time for this case was only about 100 years. We performed two sets of simulations: 

(i) The supernova event in the same cluster core that in case (2): 48 O-stars and 2 Wolf Rayets in a 2 pc radius core. The supernova explodes in the wind of one of the Wolf Rayets.  We used the simulation domain  of $9 \times 9 \times 9 ~\rm{pc}^3$ with a higher resolution in the cluster core and a total number of cells $500^3$). As in Section \ref{sec:clustprop}, particles were injected at shocks of SN and stellar winds at the same moment of time. This approach is applicable to estimate the acceleration rate and maximum energies of particles. We injected particles after $\sim$ 250 years after the explosion.
In Fig. \ref{sn_mf1} the map of the magnetic field in a cluster with SN is shown. Despite we can not take into account any of cosmic ray driven instabilities with MHD approach, we still obtain a significant magnetic field amplification -- to the hundreds of $\mu G$ near the SNR shock. Multiple colliding shock systems are formed as the remnant travels through the cluster core. They provide the additional confinement of particles in the acceleration region. 

\begin{figure}[ht]
\centering
\includegraphics [width=85mm] {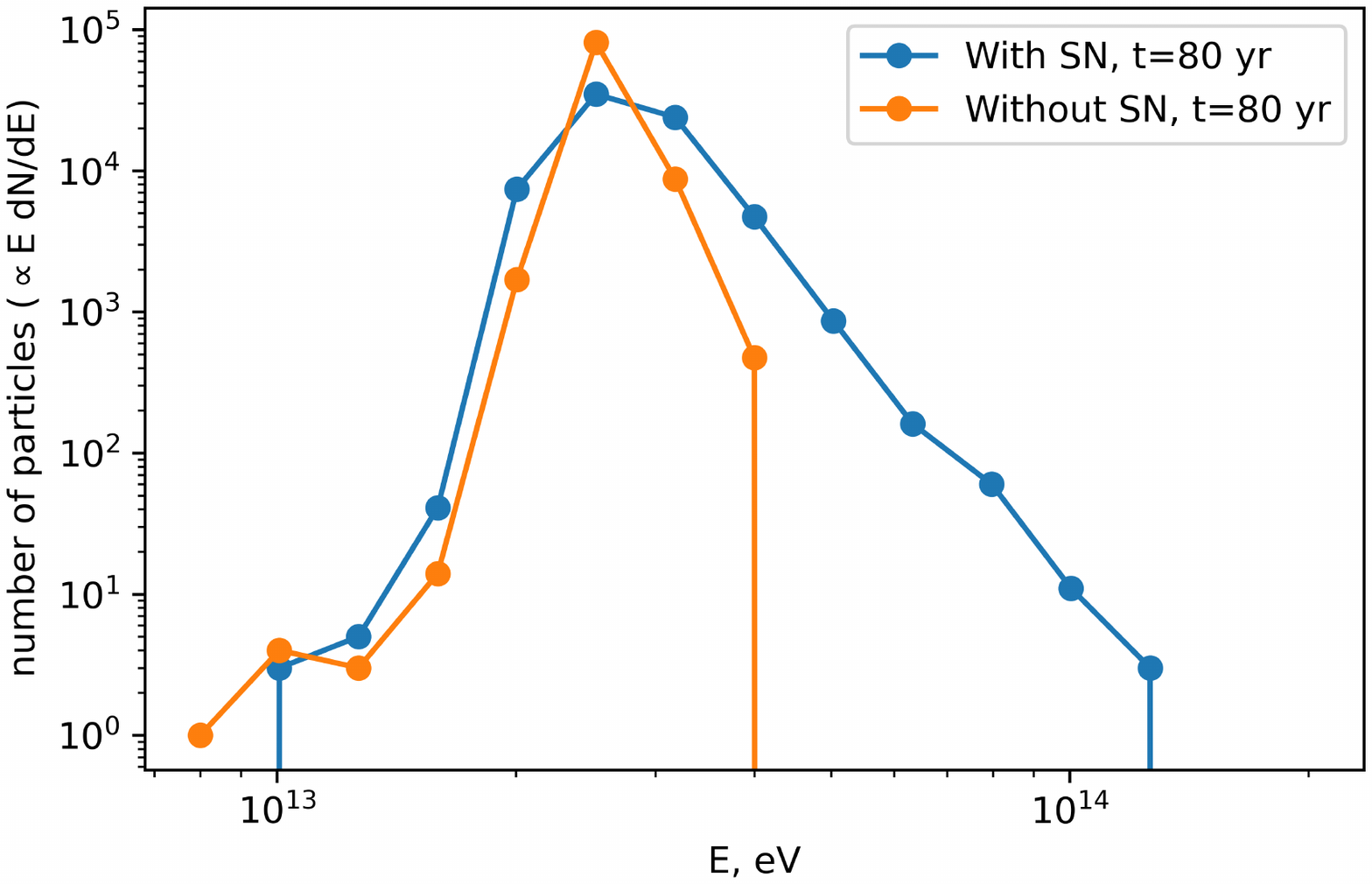}
\caption{The comparison of the particle spectra after 80 years of simulation for the clusters with and without a supernova event.}
\label{sn_nosn_spec}
\end{figure}

One can expect that given the high velocity of SN forward  shock ($\sim7000~\rm{km~s^{-1}}$ at the initialization) and highly amplified magnetic field near the shock ($\sim300~\mu G$, see Fig.~\ref{sn_mf1}), that the acceleration will be faster than in the cluster without SN. Indeed, the acceleration time can be estimated as  \citep[see][]{BEGO2015MNRAS}:
\begin{equation}
    t_{acc}\propto \frac{D}{u_w u_{sh}},
\end{equation}
where $D$ is the diffusion coefficient and $u_w$ and $u_{sh}$ are the velocities of stellar winds and SNR shock, respectively. In Fig.~\ref{sn_nosn_spec} the comparison between the particles spectra after 80 years of simulation in the case with SN and without SN is provided.  From the Figure it is clear that the particle acceleration rate is significantly higher in the cluster with SN than without it, as expected.

(ii) We also performed a simulation for a more realistic injection energy of 10 TeV, which can easily be achieved on the SNR shock, and with the highest possible resolution $0.004 ~\rm{pc~cell^{-1}}$ in order to take into account the particle scattering on the magnetic turbulence, generated by the interaction of the stellar winds. To achieve higher resolution, we used more compact cluster with core radius of 1 pc, and the simulation domain included only the core ($2 \times 2 \times 2 ~\rm{pc}^3$ with a number of cells $500^3$). As the simulation time was much shorter than the previous modeling without SN, we could use a small domain because in $\sim100$ years a significant number of particles are still in the cluster core. For the $\sim 120$ years of simulation we obtained the maximum energy of $\sim$ 200 TeV for that case. 

Modeling demonstrates that very fast and efficient acceleration takes place in the presence of SNR in a compact cluster. To obtain the accurate spectrum and possible maximum energies, prolonged simulations are needed to trace the SNR shock passage through the cluster core, which takes up to $\sim 1000$ years \citep[see][]{Bad24}. The continuous particle injection and accurate particle confinement consideration is required (the latter is shortly discussed in the next Section). This is a target for a future modeling.

\section{Particle confinement with the shell}
An important task is to investigate particle confinement within stellar clusters. As a result of the interactions between winds from massive stars and the interstellar medium or a parent molecular cloud, dense magnetized shells of various shapes and sizes can form.  By reflecting off these shells, particles can be confined within the cluster and repeatedly re-enter acceleration regions near shock fronts.

We performed an extra simulation in order to model the particle escape in the presence of a thick shell. We took as a reference the semi-elliptical shell observed near the Westerlund 2 cluster \citep[][]{Tiwari21}. Using [C II] 158 $\mu$m data from SOFIA and CO (3–2) data from APEX telescopes, the authors provide the size, density, velocity and the magnetic field of the shell. The shell has a vertical (north-south) radius of $\sim$7.5 pc and a horizontal (east–west) radius of
 $\sim$4 pc with a thickness of $\sim$1 pc. Its velocity is about 13 $\rm{km ~s^{-1}}$ and mass is $\sim 2.5 \times 10^4 \Msun$. The density of the shell reaches $4 \times 10^3 ~ \rm{cm}^{-3}$. 

 Directly simulating the formation of such a complex shell structure from the interaction of cluster winds with an asymmetric dense cloud would require a sophisticated and computationally expensive model. Therefore, we inserted an existing semi-elliptical shell near the cluster (at $\sim$ 3 pc from the cluster core of 1 pc radius) with the parameters specified in \citet[][]{Tiwari21}. The magnetic field inside the shell was set manually as 30~$\mu$G at the beginning of the calculation and remained about this magnitude up to the end. We simulated the cluster winds interactions with the shell for about $10^5$ years. After this time the cluster-driven turbulence filled the simulation domain and the shell moved to its approximate observed position. In this medium we injected $10^5$ particles in the centre of the cluster. Here we did not aim to explore the acceleration processes; we were interested only in particle propagation, so we took the injection energy of the order of maximum attainable --- 100 TeV. We followed particles for about 5000 years.

\begin{figure}[ht]
\centering
\includegraphics [width=85mm] {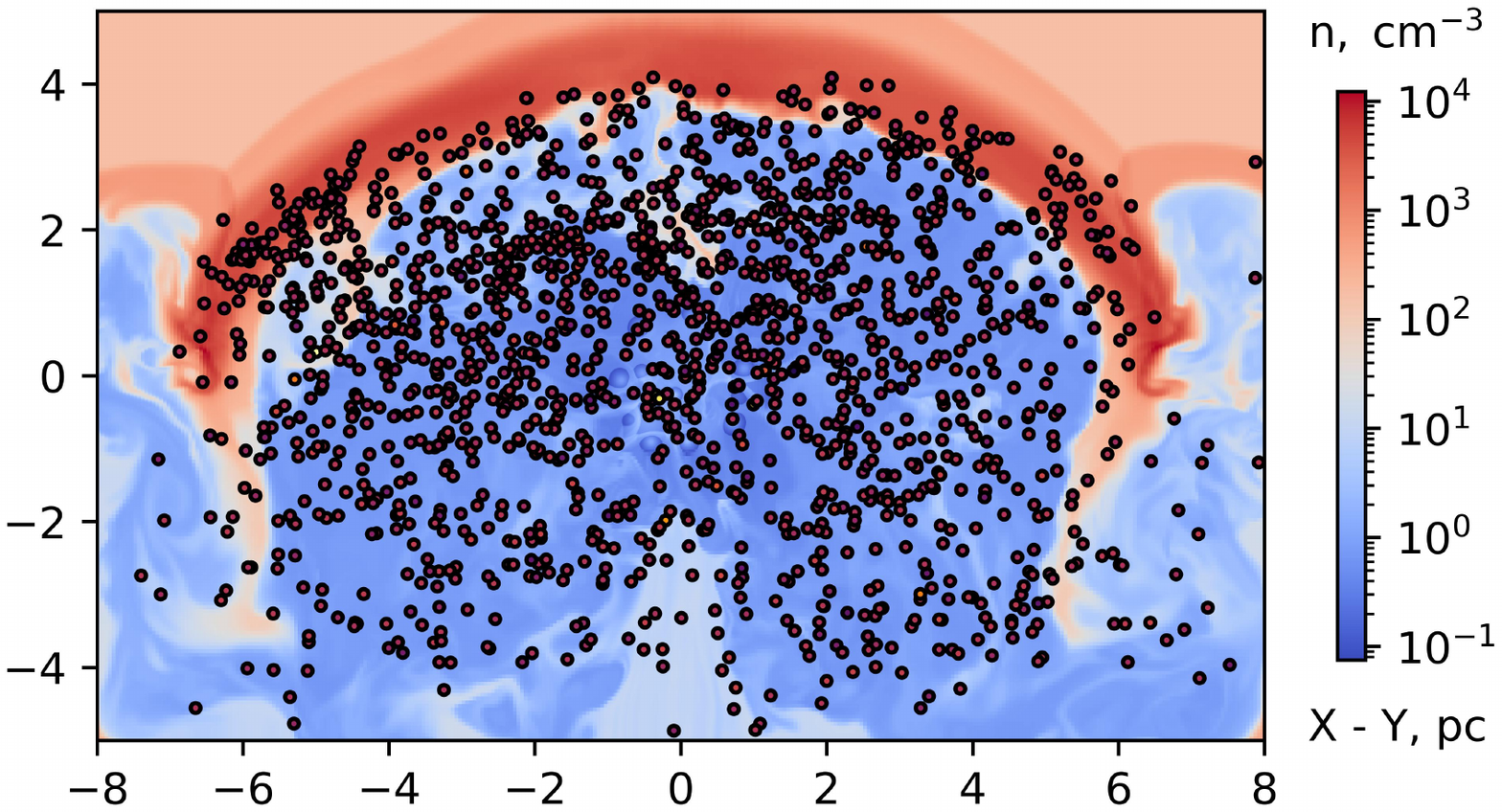}
\caption{The particles spatial distribution after 2500 years of simulation with shell. Particles do not penetrate the shell and do not escape through the upper bound.}
\label{shell}
\end{figure}

 The density map of the domain with the distribution of particles after 2500 years from the injection is shown in Fig.\ref{shell}. The red structure at the upper half of the Figure is the semi-elliptical shell. One can see that the presence of the shell leads to the suppression of particle escape from the one side, while from the other side they escape freely. Particles then reflect off the shell, accumulate in the upper half, and have a chance to return back to the cluster core (the center of the domain), and, therefore, get to one of the acceleration sites. The probabilities of particle return to the core can be extracted from the current simulation. Knowing the probabilities of particle reflection will allow us not to simulate the surroundings and simulate just the core, improving the resolution. This may be useful for the purpose of future research of the particle acceleration after the SN explosion in the compact cluster core. 

\section{Conclusions}

We present direct simulations of particle propagation and acceleration in the turbulent environments of young massive star clusters, performed using the 3D MHD PLUTO code. Our study addresses the cosmic ray acceleration in these compact stellar clusters as a complementary mechanism to standard supernova remnant shock acceleration. 

The simulations reveal significant magnetic field amplification within the cluster core, reaching magnitudes of $\sim$300 $\mu$G. This amplification arises from the interacting powerful winds of massive stars and the turbulent dynamo processes they drive. Stronger magnetic fields enhance particle confinement  and affect particle acceleration rates.

We demonstrate that protons can be accelerated to energies of $\sim$400 TeV within YMSC cores over timescales of $\sim$4000 years. The maximum energies achieved exceed those typical for isolated SNR shocks. 
%\red{Analysis of high-energy particles' trajectories reveals that diffusive shock acceleration at stellar wind termination shocks is the primary acceleration mechanism. } 
Particles gain most of their energy while confined near the shock fronts and once they diffuse away, energy gain ceases significantly even though the model does not account for the turbulence amplification by CR driven instabilities in the shock vicinity. The account for the effect requires kinetic modeling and will be discussed elsewhere. 

Modeling the expansion of a young SNR within the cluster core reveals a highly efficient acceleration process. Particles can reach energies $\sim$100–200 TeV in less than 100 years, significantly faster than the acceleration in the cluster wind environment without SN. 

An important result of the modeling is that the magnetic fields produced naturally in MHD simulations are strong enough to confine particles up to hundreds of TeV, both in cases with and without SNRs. We show that the large-scale magnetic turbulence and mirror-like structures arising in MHD model can effectively scatter and trap particles near the shock fronts. Although our modeling does not take into account cosmic ray driven instabilities, e.g. Bell and resonant instabilities, and the scale of turbulence is limited below with the resolution, we still obtained the significant cosmic ray acceleration. 
Given all that, our model can be considered 'minimal', yet it is sufficient to demonstrate that YMSCs are capable of accelerating particles up to hundreds of TeV.  To explore the limitations of 'minimal' model, we can estimate the magnetic field amplification due to the CR driven instabilities as $\langle B^2 \rangle/4 \pi \lesssim 0.1 \rho u^2$, where $u$ is the typical velocity of the flow without a fast supernova ejecta. This estimate gives us the turbulent magnetic field of about several hundreds of $\mu$G, which is of the order of the amplified field obtained in our MHD simulations. Thus, we do not expect the drastic increase in maximum energies of accelerated particles, although, particle spectra can be sensitive to the turbulence. 

Simulations with a dense, magnetized shell, reproducing a structure observed around  Westerlund 2, demonstrate its effectiveness in suppressing particle escape. Particles reflect off the shell, accumulate within the cluster volume, and have a probability of returning to the core. This confinement mechanism can be crucial for increasing the time spent by a particle near the shocks and achieving the highest energies of accelerated CRs.

While detailed spectral reconstruction requires future work with more realistic parameters, the maximum energies obtained strongly support YMSCs as viable sources of very high energy Galactic cosmic rays.

\section*{Acknowledgements}
We wish to thank the anonymous reviewers for their useful comments and suggestions. We also thank the developers of PLUTO code and the corresponding analysis tool pyPLUTO \citep{Mattia2025}. Particle acceleration simulation by M.E.K. was supported by the Foundation for the Advancement of Theoretical Physics and Mathematics “BASIS”. Particle spatial and spectral distributions were computed at the Tornado subsystem of the supercomputer center of Peter the Great St. Petersburg Polytechnic University, http://scc.spbstu.ru. 3D MHD modeling of plasma flows and magnetic fields in the compact stellar cluster by A.M.B. and D.V.B. was supported by the RSF grant 25-72-20007.

%% Bibliography
%% Author year style
\bibliographystyle{model5-names}
\biboptions{authoryear}
\bibliography{bib_DCE2}

\end{document}